\documentstyle[12pt]{article}
\newcommand\be{\begin{equation}}
\newcommand\ee{\end{equation}}
\newcommand\bea{\begin{eqnarray}}
\newcommand\eea{\end{eqnarray}}
\newcommand\ket[1]{|#1\rangle}

\newcommand\braket[2]{\langle #1|#2\rangle}
\newcommand{\fatalpha}{{\bf \alpha \kern -0.44em \alpha}}
\newcommand{\fatsigma}{{\bf \sigma \kern -0.54em \sigma}}
\newcommand{\tpchi}{{\bf \chi \kern -0.35em \chi}}
\newcommand{\llambda}{{\bf \lambda \kern -0.45em \lambda}}

\bibliography{plain}
\title{\bf Study of continuous-time quantum walks on
quotient graphs via quantum probability theory }\vspace{20mm}
\author{ S. Salimi
  \thanks{Corresponding author:  E-mail addresses: shsalimi@uok.ac.ir (S. Salimi)}
 \\ {\small Department of Physics,
University of Kurdistan, Tel-Fax: +98-871-6660075 , Sannandaj,
Iran.} \\}  \pagebreak

\vspace{20mm}
\begin{document}
\maketitle \vspace{15mm}
\newpage
\begin{abstract}
In the present paper, we study the continuous-time quantum  walk on
quotient graphs. On such graphs, there is a straightforward
reduction of problem to a subspace that can be considerably smaller
than the original one. Along the lines of reductions, by using the
idea of calculation of the probability amplitudes for
continuous-time quantum walk in terms of the spectral distribution
associated with the adjacency matrix of graphs [Jafarizadeh and
Salimi (Ann. Phys 322(2007))], we show the continuous-time quantum
walk on  original graph $\Gamma$ induces a continuous-time quantum
walk on quotient graph $\Gamma_H$. Finally, for example we
investigate continuous-time quantum walk on some quotient Cayley
graphs.


{\bf PACs Index: 03.65.Ud }
\end{abstract}

\vspace{70mm}
\newpage
\section{Introduction}
Two types of quantum walks, discrete  and continuous time, were
introduced as the quantum mechanical extension of the corresponding
random walks and have been extensively studied over the last few
years \cite{adz, Kempe}. In recent years, quantum walk has gained
considerable interest in the quantum information and computation
research areas due to its potential applications. In particular, the
study of continuous-time quantum walks (CTQW) on graphs has shown
promising applications in the algorithmic and implementation
aspects. As an alternate algorithmic technique to the Quantum
Fourier Transform and the Amplitude Amplification techniques, Childs
et al.\cite{child} demonstrated the power of CTQW algorithm for
solving a specific blackbox graph search problem. A study of quantum
walks on simple graph is well known in physics(see \cite{fls}).
Recent studies of quantum walks on more general graphs were
described in \cite{ccdfgs, fg,cfg,abnvw,aakv,mr,kem,khovi}. Some of
these works studies the problem in the important context of
algorithmic problems on graphs and suggests that quantum walks is a
promising algorithmic technique for designing future quantum
algorithms.

One approach for investigation of CTQW on graphs is using the
spectral distribution associated with the adjacency matrix of graphs
\cite{js,jsa, jsas, konno1, konno2}. Authors in Ref.\cite{js} have
introduced a new method for calculating the probability amplitudes
of quantum walk based on spectral distribution. The method of
spectral distribution only requires simple structural data of graph
and allows us to avoid a heavy combinational argument often
necessary to obtain full description of spectrum of the adjacency
matrix. In view of the fact that the quotient graphs are important
to generate the Crystallographic nets, it is tempting to try to
investigate the CTQW on quotient graphs. On such graphs, there is a
straightforward reduction of problem to a subspace that can be
considerably smaller than the original one. In this paper, along the
lines of similar reductions and the idea calculation of the
probability amplitudes for CTQW in terms of the spectral
distribution associated with the adjacency matrix of graphs
\cite{js},  we show the CTQW on graph $\Gamma$ reductions to CTQW on
quotient graph $\Gamma_H$ such that its adjacency matrix is
contained  two Szeg\"{o}- Jacobi sequences $\sqrt{\omega_i}$ and $
\alpha_i$ which denote the jumping rate from a vertex to its
neighbor and self loops, respectively. Finally, we calculate the
probability amplitude for CTQW on some quotient Cayley graphs.

 The organization of the paper is as follows. In
Section 2, we give a brief outline of Cayley graphs and automorphism
groups. In Section $3$, we investigate the action of groups and
stratification, quantum decomposition for adjacency matrix of
graphs, quotient graph, the method for obtaining spectral
distribution $\mu$ and devoted to the method of computing the
amplitude for CTQW, through spectral distribution $\mu$ of the
adjacency matrix $A$. In the Section $4$
 we calculate the  probability amplitude  for CTQW on some quotient Cayley graphs. The paper is
ended with a brief conclusion.

\section{Cayley graphs and automorphism groups}
A graph $\Gamma(V,E)$ consists of a non-empty  vertex set $V$
together with an edge set $E$ that is a subset of
$\{\{\alpha,\beta\}|\alpha, \beta\in V,\alpha\neq \beta \}$.
 Elements of $V$ and of $E$ are called \emph{vertices} and
\emph{edges}, respectively. Two vertices $\alpha, \beta\in V$ are
called adjacent if $\{\alpha,\beta\}\in E$, and in this case we
write $\alpha\sim \beta$. Let $l^2(V)$ denote the Hilbert space of
C-valued square-summable functions on
 V, and $\{\ket{\alpha}|\ \alpha\in V\}$ becomes a complete orthonormal basis of
$l^2(V)$. The adjacency matrix $A=(A_{\alpha
\beta})_{\alpha,\beta\in V}$ is defined by

\[
A_{\alpha \beta} = \left\{
\begin{array}{ll}
1 & \mbox{if $ \alpha\sim \beta$}\\
0 & \mbox{otherwise.}
\end{array}
\right.
\]

which is considered as an operator acting in $l^2(V)$ in such a way
that
$$
A\ket{\alpha}=\sum_{\alpha\sim \beta}\ket{\beta}, \;\;\;\;\alpha\in
V.
$$
Obviously, (i) $A$ is symmetric; (ii) an element of $A$ takes a
value in $\{0, 1\}$; (iii) a diagonal element of $A$ vanishes.
Conversely, for a non-empty set $V$, a graph structure is uniquely
determined by such a matrix indexed by $V$. The \emph{degree} or
\emph{valency} of a vertex $\alpha\in V$ is defined by
$$
\kappa(\alpha)=|\{\beta\in V| \alpha\sim \beta\}|,
$$
where $|.|$ denotes the cardinality.

Cayley graphs are defined in terms of a group $G$ and a set $R$ of
elements from $G$, chosen such that the identity element $e\not\in
R$. Then the Cayley graph of $G$ with respect to $R$, denoted here
by
$\Gamma(G,R)$, is defined by \cite{gross}:\\
1. Elements of $G$ are vertices of $\Gamma(G,R)$,\\
2. For $g,h \in G$, a directed edge from $g$ to $h$ exists if and
only if $hg^{-1}\in R$.\\
 Equivalently, this definition is that from
any vertex $g$ of a Cayley graph, there are $|R|$ outgoing edges,
one to each of the vertices $g \ r$, $\forall \ r\in R$. A Cayley
graph will be connected if and only if the set $R$ is a generating
set for $G$, it will be undirected if $r^{-1}\in R$,$\forall \ r\in
R$. Cayley graphs are always regular, and the degree of a Cayley
graph is $|R|$, the cardinality of the generating set.

An automorphism of a graph is a permutation $\sigma$  on the
vertices of $\Gamma$ so that for every edge $\alpha\sim \beta$ of
$\Gamma$, $\sigma(\alpha\sim \beta) =\sigma(\alpha)\sim
\sigma(\beta)$ is an edge of $\Gamma$. Thus, each automorphism of
$\Gamma$ is a one-to-one and onto mapping of the vertices of
$\Gamma$ which preserves adjacency. This implies that an
automorphism maps any vertex onto a vertex of the same degree. The
set of all such permutations is the automorphism group of the graph
which is denoted by $Aut(\Gamma)$. For example, let $\Gamma_1$ be
the graph in Fig.1, let $\sigma$ be the permutation $(13)(24)$ and
$\tau$ be the permutation $(123)$. To see that $\sigma$ is an
automorphism of $\Gamma_1$, notice that the permutation of all the
edges are indeed edges of $\Gamma_1$: \\
 $\sigma (1\sim 2)=\sigma (1)\sim\sigma (2)= 3\sim 4$\\
$\sigma (2\sim 3)=\sigma (2)\sim \sigma (3)= 4\sim 1$\\
$\sigma (3\sim 4)=\sigma (3)\sim \sigma (4)= 1\sim 2$\\
$\sigma (4\sim 1)=\sigma (4)\sim \sigma (1)= 2\sim 3$\\
$\sigma (1\sim 3)=\sigma (1)\sim \sigma (3)= 3\sim 1$.\\
Now to see that $\tau$ is not an automorphism of $\Gamma_1$ notice
that $\tau(1\sim 4) =\tau(1)\sim \tau(4)= 2 \sim 4 \not\in E$.

\section{Quotient graph}
\subsection{Action of an automorphism group and stratification}
We can think of letting the automorphism group of a graph act on the
set of vertices of the graph. In doing this, we form orbits of
vertices. Therefore, now consider  a subgroup $H$ of automorphism
group $Aut(\Gamma)$. We would like to know what kind of action this
subgroup has on the graph and hence on the Hilbert space. First, we
define what is meant by the term action \cite{khovi,rotman}.\\
\textbf{Definition:} If $X$ is a set and $G$ is a group, then $X$ is
a $G$ set if there is a function $f:G\times X\longrightarrow X$,
denoted by $f:(g, x)\longrightarrow gx$, such that:\\
1)$ex=x,\forall x\in X$,\\
2)$g(hx)=(gh)x,\ \forall g, h\in G \ \mbox{and} \ x\in X$.\\
\textbf{Definition:} If $X$ is a $G$-set an d $x\in X$, then the
$G$-orbit (or just orbit) of $x$ is
\begin{equation}\label{orbit1}
{\mathcal{O}}(x)= \{gx: g\in G\}\subset X
\end{equation}
The set of orbits of a $G$-set $X$ form a partition and the orbits
correspond to the equivalence classes under the equivalence relation
$x\equiv y$ defined by $y=gx$ for some $g\in G$. We can define the
action of the subgroup $H$ of the permutation group on the set of
basis elements $X$ of the Hilbert space $\mathcal{H}$ as the
multiplication of its matrix representation $\sigma(H)$ (in the
basis given by the vectors $X$) with a basis vector. This is a
well-defined action since $\sigma(e)\ket{x}=\ket{x}$ and
$\sigma(g)(\sigma(h)\ket{x})=(\sigma(g)\sigma(h))\ket{x}=\sigma(gh)\ket{x}$.
Therefore, the set $X$ is partitioned into orbits under the action
of $H$. Since $H$ is a subgroup of the automorphism group, these
orbits can be related to the graph $\Gamma$.

Now consider a basis vector $\ket{\alpha}$ and its $H$-orbit
${\mathcal{O}}_{\alpha}=\{\sigma(h)\ket{\alpha}; h\in H\}$.
Therefore, due to definition of  this orbits, the graph is
stratified into a disjoint union of orbits:
\begin{equation}\label{v1}
V=\bigcup_{\alpha\in \Gamma}\mathcal{O}_\alpha.
\end{equation}
With each orbit $\mathcal{O}_\alpha$ we associate a unit vector in
$l^2(V)$ defined by
\begin{equation}
\ket{\phi_{\alpha}}=\frac{1}{\sqrt{|\mathcal{O}_\alpha|}}\sum_{h\in
H}\sigma(h)\ket{\alpha},
\end{equation}
Each of these vectors $\ket{\phi_{\alpha}}$ are orthonormal, since
they are formed from orbits and distinct orbits do not intersect and
they span the simultaneous eigenspace of eigenvalue $1$ of the
matrices $\sigma(H)$ where we denote by $\Lambda(\Gamma)$. Since
$\{\ket{\phi_{\alpha}}\}$ becomes a complete orthonormal basis of
$\Lambda(\Gamma)$, we often write
\begin{equation}
\Lambda(\Gamma)=\sum_{\alpha}\oplus \textbf{C}\ket{\phi_{\alpha}}.
\end{equation}
Hereafter, we replace the indices of the orbits with integers, i.e.,
we indicate $\mathcal{O}_\alpha$ and $\ket{\phi_{\alpha}}$ with
${\mathcal{O}}_i$ and $\ket{\phi_{i}}$, respectively ($0\leq i\leq
d$, where $d$ is the numbers of orbits ).

Let $A$ be the adjacency matrix of a graph $\Gamma$. According to
the stratification (\ref{v1}), one can obtain a quantum
decomposition for the adjacency matrices of this type of graphs as
\begin{equation}
A=A^{+}+A^{-}+A^0.
\end{equation}
where three matrices $A^+$, $ A^-$ and $A^0$  are defined as
follows: for $\alpha\in {\mathcal{O}}_i$
\[
(A^+)_{\beta\alpha} = \left\{
\begin{array}{ll}
A_{\beta\alpha} & \mbox{if $ \beta\in {\mathcal{O}}_{i+1}$}\\
0 & \mbox{otherwise.}
\end{array}
\right.
\]
\[
(A^-)_{\beta\alpha} = \left\{
\begin{array}{ll}
A_{\beta\alpha} & \mbox{if $ \beta\in {\mathcal{O}}_{i-1}$}\\
0 & \mbox{otherwise.}
\end{array}
\right.
\]
\[
(A^0)_{\beta\alpha} = \left\{
\begin{array}{ll}
A_{\beta\alpha} & \mbox{if $ \beta\in {\mathcal{O}}_i$}\\
0 & \mbox{otherwise.}
\end{array}
\right.
\]
Or, equivalently, for $\ket{\alpha}\in {\mathcal{O}}_i$,
\begin{equation}\label{qd}
A^{+}\ket{\alpha}=\sum_{\beta\in {\mathcal{O}}_{i+1}}\ket{\beta},
\;\;\;\;\ A^{-}\ket {\alpha}=\sum_{\beta\in
{\mathcal{O}}_{i-1}}\ket{\beta}, \;\;\;\;\
A^{0}\ket{\alpha}=\sum_{\beta\in {\mathcal{O}}_i}\ket{\beta},
\;\;\;\;\
 \end{equation}
 for $\alpha\sim \beta$. By lemma 2.2, \cite{nob} if $G$ is
invariant under the quantum components $A^\varepsilon$,
$\varepsilon\in \{+,-,0\}$,  then there exist two Szeg\"{o}- Jacobi
sequences $\{\omega_i\}_{i=1}^{\infty}$ and
$\{\alpha_i\}_{i=1}^{\infty}$ derived from $A$, such that
\begin{equation}\label{v5}
A^{+}\ket{\phi_{i}}=\sqrt{\omega_{i+1}}\ket{\phi_{i+1}}, \;\;\;\
i\geq 0
\end{equation}
\begin{equation}\label{v6}
A^{-}\ket{\phi_{0}}=0, \;\;\
A^{-}\ket{\phi_{i}}=\sqrt{\omega_{i}}\ket{\phi_{i-1}}, \;\;\;\ i\geq
1
\end{equation}
\begin{equation}\label{v7}
A^{0}\ket{\phi_{i}}=\alpha_{i+1}\ket{\phi_{i}}, \;\;\;\ i\geq 0,
\end{equation}
where
$\sqrt{\omega_{i}}=\frac{|{\mathcal{O}}_{i+1}|^{1/2}}{|{\mathcal{O}}_{i}|^{1/2}}\kappa_{-(\beta)}$,
$\kappa_{-(\beta)}=|\{\alpha\in {\mathcal{O}}_{i}| \alpha\sim
\beta\}|$ for $\beta\in {\mathcal{O}}_{i+1}$ and
$\alpha_{i+1}=\kappa_{0(\beta)}$, such that
$\kappa_{0(\beta)}=|\{\alpha\in {\mathcal{O}}_{i}| \alpha\sim
\beta\}|$ for $\beta\in {\mathcal{O}}_{i}$. In particular
$(\Lambda(\Gamma), A^+, A^-)$ is an interacting Fock space
associated with a Szeg\"{o}-Jacobi sequence $\{\omega_i\}$.
\subsection{Quotient graphs, spectral distribution of the adjacency matrix graph and CTQW}
Based on the action on a graph $\Gamma$ of the subgroup $H$ of its
automorphism group, the quotient graph  $\Gamma_H$ ( $\Gamma/H$) is
that graph whose vertices are the $H$-orbits, and two such vertices
${\mathcal{O}}_{i}$ and ${\mathcal{O}}_{j}$ are adjacent in
$\Gamma_H$ if and only if there is an edge in $\Gamma$ joining a
vertex of ${\mathcal{O}}_{i}$ to a vertex of ${\mathcal{O}}_{j}$.

With due attention to the idea of calculation of the probability
amplitudes for continuous-time quantum walk, in terms of the
spectral distribution associated with the adjacency matrix of graphs
\cite{js}, CTQW on graph $\Gamma$ induces a CTQW on quotient graph
$\Gamma_H$ as long as adjacency matrix is as
\begin{equation}
A_{\Gamma_H}=\left(
\begin{array}{ccccc}
 \alpha_1 & \sqrt{\omega_1} & 0 & ... &... \\
      \sqrt{\omega_1} & \alpha_2 & \sqrt{\omega_2} & 0 &... \\
      0 & \sqrt{\omega_2} & \alpha_3 & \sqrt{\omega_3} & ... \\
     ... & ... &...& ... &... \\
     ...& ... &0 & \sqrt{\omega_d} & \alpha_{d}\\
\end{array}
\right),
\end{equation}
where $\sqrt{\omega_i}$ and $ \alpha_i$ are the jumping rate from a
vertex to its neighbor and self loops, respectively.

 The spectral
properties of the adjacency matrix of a graph play an important role
in many branches of mathematics and physics . The spectral
distribution can be generalized in various ways. In this work,
following Refs.\cite{js, nob}, we consider the spectral distribution
$\mu$ of the adjacency matrix $A$:
\begin{equation}\label{v2}
\langle A^m\rangle=\int_{R}x^{m}\mu(dx), \;\;\;\;\ m=0,1,2,...
\end{equation}
where $\langle,\rangle$ is the mean value with respect to the
state $\ket{\phi_0}$. By condition  of QD graphs  the ''moment''
sequence $\{\langle A^m\rangle\}_{m=0}^{\infty}$ is
well-defined\cite{js,nob}. Then the existence of a spectral
distribution satisfying (\ref{v2}) is a consequence of Hamburger's
theorem, see e.g., Shohat and Tamarkin
[\cite{st}, Theorem 1.2].\\
 We may apply the canonical isomorphism from
the interacting Fock space onto the closed linear span of the
orthogonal polynomials determined by the Szeg\"{o}-Jacobi
sequences $(\{\omega_i\},\{\alpha_i\})$. More precisely, the
spectral distribution $\mu$ under question is characterized by
the property of orthogonalizing the polynomials $\{P_n\}$ defined
recurrently by
$$ P_0(x)=1, \;\;\;\;\;\
P_1(x)=x-\alpha_1,$$
\begin{equation}\label{op}
xP_n(x)=P_{n+1}(x)+\alpha_{n+1}P_n(x)+\omega_nP_{n-1}(x),
\end{equation}
for $n\geq 1$.

As it is shown in \cite{tsc}, the spectral distribution ì can be
determined by the following identity:
\begin{equation}\label{v3}
G_{\mu}(z)=\int_{R}\frac{\mu(dx)}{z-x}=\frac{1}{z-\alpha_1-\frac{\omega_1}{z-\alpha_2-\frac{\omega_2}
{z-\alpha_3-\frac{\omega_3}{z-\alpha_4-\cdots}}}}=\frac{Q_{n-1}^{(1)}(z)}{P_{n}(z)}=\sum_{l=1}^{n}
\frac{A_l}{z-x_l},
\end{equation}
where $G_{\mu}(x)$ is called the Stieltjes transform and $A_l$ is
the coefficient in the Gauss quadrature formula corresponding to
the roots $x_l$ of polynomial $P_{n}(x)$ and where the polynomials
$\{Q_{n}^{(1)}\}$ are defined
recurrently as\\
        $Q_{0}^{(1)}(x)=1$,\\
    $Q_{1}^{(1)}(x)=x-\alpha_2$,\\
    $xQ_{n}^{(1)}(x)=Q_{n+1}^{(1)}(x)+\alpha_{n+2}Q_{n}^{(1)}(x)+\omega_{n+1}Q_{n-1}^{(1)}(x)$,\\
    for $n\geq 1$.

Now if $G_{\mu}(z)$ is known, then the spectral distribution $\mu$
can be recovered from $G_{\mu}(z)$ by means of the Stieltjes
inversion formula:
\begin{equation}\label{m1}
\mu(y)-\mu(x)=-\frac{1}{\pi}\lim_{v\longrightarrow
0^+}\int_{x}^{y}Im\{G_{\mu}(u+iv)\}du.
\end{equation}
Substituting the right hand side of (\ref{v3}) in (\ref{m1}), the
spectral distribution can be determined in terms of $x_l,
l=1,2,...$, the roots of the polynomial $P_n(x)$, and  Guass
quadrature constant $A_l, l=1,2,... $ as
\begin{equation}\label{m}
\mu=\sum_l A_l\delta(x-x_l)
\end{equation}
 ( for more details see Refs. \cite{js,jsa,tsc,st}.)

Due to using the quantum decomposition relations (\ref{v5},
\ref{v6},\ref{v7}) and the recursion relation (\ref{op}) of
polynomial $P_n(x)$, the other matrix elements $\label{cw1}
\braket{\phi_{k}}{A^m\mid \phi_0}$ can be written as
\begin{equation}\label{cw1}
\braket{\phi_{m}}{A^l\mid
\phi_0}=\frac{1}{\sqrt{\omega_1\omega_2\cdots \omega_{m}
}}\int_{R}x^{l}P_{m}(x)\mu(dx),  \;\;\;\;\ l=0,1,2,....
\end{equation}
where for obtaining of amplitudes of CTQW in terms of spectral
distribution associated with the adjacency matrix of graphs is
useful \cite{js}.

CTQW  on graph were introduced as the quantum mechanical analogue of
classical its, which are defined by replacing Kolmogorov's equation
(master equation) of continuous-time classical random walk on a
graph with Schr\"{o}dinger's equation. A state $\ket{\phi_0}$
evolves in time as $\ket{\phi(t)} = U(t)\ket{\phi_0}$, where
$U(t)=e^{-iHt}$ is the quantum mechanical time evolution operator
(we have set $m = 1$ and $\hbar = 1$). It is natural to choose the
Laplacian of the graph, defined as $L=A-D$ as Hamiltonian of walk,
where $D$ is a diagonal matrix with entries $D_{jj}=deg(\alpha_j)$.
On $d$-regular graphs, $D=\frac{1}{d}I$ and since $A$ and $D$
commute, we get
\begin{equation}
e^{-itH} = e^{-it(A-\frac{1}{d} I)} = e^{-it/d}e^{-itA}.
\end{equation}
Hence we can consider $H = A$. Therefore for obtaining the
probability amplitude of CTQW at orbit $m$ at time $t$ can be
replaced time evolution operator with operator $A^l$ in the equation
(\ref{cw1}) as
\begin{equation}\label{v4}
q_{m}(t)=\braket{\phi_{m}}{e^{-iAt}\mid
\phi_0}=\frac{1}{\sqrt{\omega_1\omega_2\cdots\omega_{m}}}\int_{R}e^{-ixt}P_{m}(x)\mu(dx).
\end{equation}
The conservation of probability $\sum_{m=0}{\mid q_{m}(t)\mid}^2=1$
follows immediately from Eq.(\ref{v4}) by using the completeness
relation of orthogonal polynomials $P_n(x)$. Obviously evaluation of
$q_{m}(t)$ leads to the determination of the amplitudes at sites
belonging to the stratum $V_m$, as it is proved in the appendix A
\cite{js}, the walker has the same amplitude at the vertex belonging
to the same stratum, i.e., we have $q_{im}(t)=\frac{q_{m}(t)}{\mid
V_m\mid}, i=0,1,...,\mid V_m\mid$, where $q_{im}(t)$ denotes the
amplitude of the walker at $i$th vertex of $m$th stratum.

Formula (\ref{v4}) indicates a canonical isomorphism between the
interacting Fock space CTQW on QD graphs and the closed linear
span of the orthogonal polynomials generated by recursion
relations (\ref{op}). This isomorphism was meant to be, a
reformulation of CTQW (on QD graphs), which describes quantum
states by polynomials (describing quantum state $\ket{\phi_m}$ by
$P_m(x)$), and  make a correspondence between functions of
operators ($q$-numbers) and functions of classical quantity
($c$-numbers), such as the correspondence between $e^{-iAt}$ and
$e^{-ixt}$. This isomorphism is similar to the isomorphism
between Fock space of annihilation and creation operators $a$,
$a^{\dag}$ with space of functions of coherent states parameters
in quantum optics, or the isomorphism between Hilbert space of
momentum and position operators, and spaces of function defined
on phase space in Wigner function formalism.

At the end, by using relation of spectral distribution (\ref{m})
for finite graphs, amplitude of probability (\ref{v4}) is
agreeable with
\begin{equation}\label{fin}
q_{m}(t)=\frac{1}{\sqrt{\omega_1\omega_2\cdots\omega_{m}}}\sum_{l}A_le^{-ix_lt}P_{m}(x_l),
\end{equation}
where by straightforward  calculation one can  evaluate  the
average probability for the finite graphs as
\begin{equation}
\bar{P}(m)=\lim_{T\rightarrow \infty}\frac{1}{T}\int_{0}^{T}\mid
q_{m}(t)\mid
^2dt=\frac{1}{\omega_1\omega_2\cdots\omega_{m}}\sum_{l}A_l^2P_{m}^2(x_l).
\end{equation}
\section{Examples of quotient graphs}
In this section we provide some examples of finite quotient graphs
and use the spectral distribution to calculate the relevant
amplitudes of continuous-time quantum walks on these graphs.

\textbf{Example 1.} As the first example, consider the Cayley graph
$\Gamma(S_3, R)$ where $R=\{r_1, r_2\}=\{(1,2),(2,3)\}$. The basis
vectors of Hilbert space of walk are $\{\ket{e}, \ket{r_1},
\ket{r_2},\ket{r_1r_2}, \ket{r_2r_1}, \ket{r_2r_1r_2}\}$, where the
original and the quotient graphs are shown in Fig.2. The
automorphism group of this graph is $Aut(\Gamma(S_3, R))\simeq
R(S_3)$. Consider the subgroup $H$ which corresponds to
interchanging the vertices $r_1\longleftrightarrow r_2$ and
$r_1r_2\longleftrightarrow r_1r_2$. Therefore the orbits and unit
vectors under the action of this subgroup are,
$$
{\mathcal{O}}_0=\{\ket{e}\},  \;\;\;\ \ket{\phi_0}=\ket{e}
$$
$$
{\mathcal{O}}_1=\{\ket{r_1}, \ket{r_2}\},  \;\;\;\
\ket{\phi_1}=\frac{1}{\sqrt{2}}(\ket{r_1}+\ket{r_2})
$$
$$
{\mathcal{O}}_2=\{\ket{r_1r_2}, \ket{r_2r_1}\},  \;\;\;\
\ket{\phi_2}=\frac{1}{\sqrt{2}}(\ket{r_1r_2}+\ket{r_2r_1})
$$
\begin{equation}
{\mathcal{O}}_3=\{\ket{r_2r_1r_1}\},  \;\;\;\
\ket{\phi_3}=\ket{r_2r_1r_1}.
\end{equation}
In this case the  two Szeg\"{o}- Jacobi sequences $\{\omega_i\}$
and $\{\alpha_i\}$ are given by
\begin{equation}
\omega_1=2, \;\ \omega_2=1, \;\ \omega_3=2, \;\
\alpha_1=\alpha_2=\cdots=0.
\end{equation}
Hence, The Stieltjes transform and spectral distribution are
obtained as
\begin{equation}
G_{\mu}(z)={\frac {  {z}^{3}-3z  }{{z}^{4}-5{z}^{2}+4}}, \;\;\;\;\
\mu(x)=\frac{1}{3}(\delta(x-1)+\delta(x+1))+\frac{1}{6}(\delta(x-2)+\delta(x+2)).
\end{equation}
Using equations(\ref{v4}) or (\ref{fin}), we can calculate the
probability amplitude of orbits as
$$
q_{0}(t)=\frac{1}{3}(\cos(2t)+2\cos(t)),
$$
$$
q_{1}(t)=\frac{-2i}{3\sqrt{2}}(\sin(2t)+\sin(t)),
$$
$$
q_{2}(t)=\frac{2}{3\sqrt{2}}(\cos(2t)-\cos(t)),
$$
\begin{equation}
q_{3}(t)=\frac{i}{3}(-2\sin(2t)+\sin(t)).
\end{equation}
\textbf{Example 2.} In the second example we consider the Cayley
graph $\Gamma(S_3, R)$ where $R=\{r_1, r_2, r_3\}=\{(1,2),(2,3),
(1,3)\}$. The original and the quotient graphs are shown in Fig.3.
In this case we consider the subgroup $H$ of its automorphism group
which corresponds to interchanging the vertices
$r_1\longleftrightarrow r_2\longleftrightarrow r_3$ and
$r_1r_2\longleftrightarrow r_1r_2$. The orbits and unit vectors
under the action of this subgroup are,
$$
{\mathcal{O}}_0=\{\ket{e}\},  \;\;\;\ \ket{\phi_0}=\ket{e}
$$
$$
{\mathcal{O}}_1=\{\ket{r_1}, \ket{r_2}, \ket{r_3}\},  \;\;\;\
\ket{\phi_1}=\frac{1}{\sqrt{3}}(\ket{r_1}+\ket{r_2}+\ket{r_3}),
$$
\begin{equation}
{\mathcal{O}}_2=\{\ket{r_1r_2}, \ket{r_2r_1}\},  \;\;\;\
\ket{\phi_2}=\frac{1}{\sqrt{2}}(\ket{r_1r_2}+\ket{r_2r_1}).
\end{equation}
Therefore  the  two Szeg\"{o}- Jacobi sequences $\{\omega_i\}$,
$\{\alpha_i\}$, the Stieltjes transform and spectral distribution
are given by
$$
\omega_1=3, \;\ \omega_2=6, \;\;\ \alpha_1=\alpha_2=\cdots=0.
$$
\begin{equation}
G_{\mu}(z)={\frac {  {z}^{2}-6  }{{z}^{3}-9{z}}}, \;\;\;\;\
\mu(x)=\frac{2}{3}\delta(x)+\frac{1}{6}(\delta(x-3)+\delta(x+3)).
\end{equation}
Hence by using equations(\ref{v4}) or (\ref{fin}), one can calculate
the probability amplitude of orbits as
$$
q_{0}(t)=\frac{1}{3}(\cos(3t)+2),
$$
$$
q_{1}(t)=\frac{-i}{\sqrt{3}}\sin(3t),
$$
\begin{equation}
q_{2}(t)=\frac{2}{3\sqrt{2}}(\cos(3t)-1).
\end{equation}
\textbf{Example 3.} In this example we consider the Cayley graph
$\Gamma({\mathcal{Z}}_n, R)$ with  $R=\{1, n-1\}$ which is well
known as the cycle graph $C_n$. The original and quotient graphs are
shown in Fig.4. Consider the subgroup $H$ of its automorphism
correspond to interchanging the vertices $1\longleftrightarrow n-1,
\ 2\longleftrightarrow n-2,\ ... \ k\longleftrightarrow n-k, ...$.
The orbits and unit vectors under the action of this subgroup are
$$
{\mathcal{O}}_0=\{\ket{e}\},  \;\;\;\ \ket{\phi_0}=\ket{e}
$$
$$
{\mathcal{O}}_1=\{\ket{1}, \ket{n-1}\},  \;\;\;\
\ket{\phi_1}=\frac{1}{\sqrt{2}}(\ket{1}+\ket{n-1}),
$$
$$ \vdots $$
\begin{equation}
{\mathcal{O}}_{\frac{n-1}{2}}=\{\ket{\frac{n-1}{2}},
\ket{\frac{n+1}{2}}\}, \;\;\;\
\ket{\phi_{\frac{n-1}{2}}}=\frac{1}{\sqrt{2}}(\ket{\frac{n-1}{2}}+\ket{\frac{n+1}{2}}),
\end{equation}
for the $n$ odd, but if the $n$ is even therefor the last orbit and
unit vector is given by
\begin{equation}
{\mathcal{O}}_{\frac{n}{2}}=\{\ket{\frac{n}{2}}\}, \;\;\;\
\ket{\phi_{\frac{n}{2}}}=\ket{\frac{n}{2}}.
\end{equation}
Therefore the two Szeg\"{o}- Jacobi sequences $\{\omega_i\}$,
$\{\alpha_i\}$ and spectral distribution
 are given by:\\
 if $n$ is  odd,
 $$
\omega_1=2, \;\ \omega_2=\omega_3=\cdots=\omega_{\frac{n-1}{2}}=1,
\;\;\ \alpha_1=\alpha_2=\cdots=0, \;\ \alpha_{\frac{n+1}{2}}=1.
$$
\begin{equation}\label{d1}
\mu=\frac{1}{n}(\delta(x-2)+2\sum_{l=1}^{\frac{n-1}{2}}\delta(x-2\cos(\frac{2l\pi}{n}))),
\end{equation}
if $n$ is even,
$$
  \omega_1=2 ,\;\;\;\; \omega_2=\omega_3=\cdots\omega_{\frac{n}{2}-1}=1 ; \;\;\
  \omega_{\frac{n}{2}}=2;   \;\;\;\;\;\
  \alpha_1=\alpha_2=\cdots=0.
$$
\begin{equation}
\mu=\frac{1}{n}(\delta(x-2)+\delta(x+2))+\frac{2}{n}\sum_{l=1}^{n/2-1}\delta(x-2\cos(\frac{2l\pi}{n})).
\end{equation}
Then by using equations(\ref{v4}) or (\ref{fin}),  one can calculate
the probability amplitude of orbits which as example we obtain for
$0$-th orbit as:\\
if $n$ is odd,
\begin{equation}
q_0(t)=\frac{1}{n}(e^{-it}+2\sum_{l=1}^{\frac{n-1}{2}}e^{-it\cos{2l\pi/n}}).
\end{equation}
 if $n$ is even,
\begin{equation}
q_0(t)=\frac{2}{n}(\cos{t}+\sum_{l=1}^{n/2-1}e^{-it\cos{2l\pi/n}}),
\end{equation}
where the results are in agreement with those of Ref. \cite{js}. In
the limit of the large $n$, the quotient graph is infinite line
graph $\mathcal{Z}$.

\section{Conclusion}
By using the method of calculation of the probability amplitude for
CTQW on graph\cite{js}, we have shown CTQW on graph $\Gamma$ induces
a CTQW on quotient graph  $\Gamma_H$. Then we obtained the
probability amplitude of  CTQW on some quotient Cayley graphs. In
view of the fact that the quotient graphs are important to generate
Crystallographic nets, it is possible to generalize this method for
investigating CTQW on Crystallographic nets, which is under
investigation.

\newpage
{\bf Figure Captions}

{\bf Figure-1:} The graph $\Gamma_1$.

{\bf Figure-2:} The graph $\Gamma(S_3, \{(1,2),(2,3)\})$ and its
quotient graph.

{\bf Figure-3:} The graph $\Gamma(S_3, \{(1,2),(2,3),(1,3)\})$ and
its quotient graph.

{\bf Figure-4:} The graph $\Gamma({\mathcal{Z}}_5, \{1,4\})$ and its
quotient graph

\end{document}